\documentclass[twocolumn]{aastex62}
\usepackage{graphics,epsf}
\usepackage{amsmath}                % American Mathematical Society package
\usepackage{amsfonts}               % American Mathematical Society fonts
\usepackage{amssymb}                % American Mathematical Society symbol
\usepackage{epsfig}                 % EPS figures
\usepackage{graphicx}
\usepackage{float}
\usepackage{color}
\usepackage[para,online,flushleft]{threeparttable}

\newcommand{\cm}{{~\rm cm}}
\newcommand{\km}{{~\rm km}}
\newcommand{\s}{{~\rm s}}

\newcommand{\g}{{~\rm g}}

\newcommand{\K}{{~\rm K}}
\newcommand{\erg}{{~\rm erg}}
\newcommand{\yr}{{~\rm yr}}

\newcommand{\AU}{{~\rm AU}}

\begin{document}

\title{Intermediate Luminosity Optical Transients (ILOTs) from merging giants}

\correspondingauthor{Efrat Sabach; Noam Soker}
\email{efrats@physics.technion.ac.il; soker@physics.technion.ac.il}

\author{Ran Segev}
\affiliation{Department of Physics, Technion, Haifa, 3200003, Israel}

\author{Efrat Sabach}
\affiliation{Department of Physics, Technion, Haifa, 3200003, Israel}

\author[0000-0003-0375-8987]{Noam Soker}
%\affil{Departmeמt of Physics, Technion, Haifa 3200003, Israel}
\affiliation{Department of Physics, Technion, Haifa, 3200003, Israel}
\affiliation{Guangdong Technion Israel Institute of Technology, Shantou 515069, Guangdong Province, China}

%%% \author{Ran Segev\altaffilmark{1}, Efrat Sabach\altaffilmark{1}, Noam Soker\altaffilmark{1,2}}

%%% \altaffiltext{1}{Department of Physics, Technion -- Israel Institute of Technology, Haifa 32000, Israel; efrats@physics.technion.ac.il; soker@physics.technion.ac.il}
%%% \altaffiltext{2}{Guangdong Technion Israel Institute of Technology, Shantou, Guangdong Province 515069, China}

\begin{abstract}
We suggest and study the formation of intermediate luminosity optical transients (ILOTs) from the merger of two cool giant stars. For the two stars to merge when both are in their giant phases the stars must have close masses at their zero age main sequence, and the orbital separation must be in the right range.
After the two giants merge, the two cores spiral-in toward each other within a common envelope. We study the energy sources of radiation in this process that include the ejection of mass that powers radiation by both recombination and by collision with previously ejected mass. This process includes no jets, unlike many other types of ILOTs, hence the event will not form a bipolar nebula.  
Using the stellar evolution numerical code \textsc{mesa} for two binary systems with stellar masses of $(15 M_\odot, 15.75 M_\odot)$ and $(31 M_\odot, 31.5 M_\odot)$, we find that the merger of the two cores releases gravitational energy that marginally ejects the entire common envelope.
This implies that in many cases the two cores merge, i.e., a fatal common envelope evolution, leading to a somewhat more luminous ILOT.
A typical ILOT from merger of two cool giant stars lasts for several months to several years and has a typical average luminosity of  $L_{\rm ILOT} \approx 10^6 (M_{\rm CE} / 10 M_\odot) L_\odot$, where $M_{\rm CE}$ is the ejected common envelope mass.
The merger-driven massive outflow forms dust, hence leading to a very red ILOT, possibly even infrared luminous and undetectable in the visible.
\end{abstract}


% ==========================================================
\section{INTRODUCTION}
 \label{sec:intro}
% ==========================================================

Observations over the years have filled-up the domain of eruptive stars with peak luminosities above typical nova luminosities but below typical supernova luminosities (e.g. \citealt{Mouldetal1990, Rau2007, Ofek2008, Prieto2009, Botticella2009, Smithetal2009, KulkarniKasliwal2009, Mason2010, Pastorello2010, Kasliwal2011, Tylendaetal2013, Kasliwal2013, Blagorodnovaetal2017, Kaminskietal2018, Pastorelloetal2018, BoianGroh2019}). Thermonuclear outbursts and explosions power some of these, while gravitational energy powers others.

We will refer to transient events that are not supernovae (SNe) and that are powered by gravitational energy as Intermediate Luminosity Optical Transients (ILOTs; \citealt{Berger2009b, KashiSoker2016, Muthukrishnaetal2019}). The heterogeneous class of ILOTs contains several subclasses. \cite{KashiSoker2016} list the following classes\footnote{See \url{http://physics.technion.ac.il/~ILOT/} for an updated list.}.
%\begin{enumerate}
%\item
(i) Intermediate-Luminous Red Transients (ILRT). These are ILOTs of evolved stars, such as asymptotic giant branch (AGB) or extreme-AGB (ExAGB) stars, like SN~2008S \citep{ArbourBoles2008} and NGC~300~OT2008-1 (\citealt{Monard2008, Bond2009}).
%\item
(ii) Giant eruptions of luminous blue variables (LBV), such as the Great Eruption of $\eta$ Carinae in the years 1837--1856, and SN Impostors, such as the pre-explosion outbursts of SN~2009ip.
%\item
(iii) Luminous Red Novae (LRN) or Red Transients (RT) or Merger-Bursts, such as V838 Mon and V1309~Sco. A full merger of two stars powers these events. Merger events of stars with sub-stellar objects also belong to this class.
%\end{enumerate}

We note that there are alternative names to ILOTs in use.
\cite{Jencsonetal2019}, as an example, do not use the name ILOT, but rather use intermediate luminosity red transients for explosions of extreme asymptotic giant branch stars, and luminous red novae for merging stars.
\cite{Pastorelloetal2019} do use the term ILOTs, but their division of ILOTs to different classes is different than that of \cite{KashiSoker2016} that we use here. 

We take the view that binary interaction powers ILOTs (e.g., \citealt{Kashietal2010, KashiSoker2010b, SokerKashi2013, McleySoker2014, Pejchaetal2016a, Pejchaetal2016b, MacLeodetal2018, Michaelisetal2018, Pastorelloetal2019}; see \citealt{Soker2016NewA} for a review).
The merger of two stars to form an ILOT can result in the destruction of one of them or the formation of a common envelope. \cite{SokerTylenda2003} and \cite{TylendaSoker2006} suggested that the ILOT V838~Mon resulted from the merger of two stars where the low mass star had been destroyed onto the more massive star. On the other hand, \cite{RetterMarom2003} and \cite{Retteretal2006} suggested an alternative model that is based on a common envelope evolution (CEE) where planets entered the envelope of the stellar progenitor of V838~Mon.
Other researchers followed and suggested the CEE scenario, but with stellar companions, to explain other ILOTs, e.g., OGLE-2002-BLG-360 \citep{Tylendaetal2013}, {V1309~Sco} \citep{Tylendaetal2011, Ivanovaetal2013a, Nandezetal2014, Kaminskietal2015}, and M31LRN~2015 \citep{MacLeodetal2017}.

In the above list of scenarios the basic energy source is high accretion rate of mass, either by a mass transfer process or by the destruction of one star. In many cases the mass transfer involves the launching of jets. This powering source is termed the high-accretion-powered ILOTs (HAPI) model \citep{KashiSoker2016}. In the present paper we study a specific type of CEE ILOT, where two giant stars merge to form a CEE. In some cases the two cores of the giant stars merge as well.

There are works that mention the process of the merger of a companion with the core of a giant, but these concentrate on a companion that is a substellar object (e.g., \citealt{HarpazSoker1994, SiessLivio1999}), a main sequence star, e.g., as in the progenitor of SN 1987A (e.g., \citealt{ChevalierSoker1989, Podsiadlowskietal1990, MenonHeger2017, Menonetal2018, Urushibataetal2018}) or unusual nucleosynthesis \citep{IvanovaPodsiadlowski2002}, a white dwarf (as in the progenitor of type Ia supernovae in the core degenerate scenario; e.g., \citealt{IlkovSoker2012}), or a neutron star (as in common envelope jet supernovae; \citealt{SokerGilkis2018, Sokeretal2019}). \cite{Soker2019Fatal} summarises these different cases and their properties.

In this paper we study the a rare type of ILOTs where the two merging stars are giants and the two cores merge.
There are other studies of the merger of two giant stars that examine other aspects. \cite{VignaGomezetal2018}, for example, examine the formation of double neutron star systems. They do not consider the merger of the two cores and the formation of an ILOT.
\cite{VignaGomezetal2019} consider the merger of two massive stars ($ > 30 M_\odot$) after they have both developed a hydrogen-exhausted core, as a route to form progenitors of pair instability supernovae. They did not study the ILOT that might result from the merger itself.
Therefore, our present study contains novel aspects of merging giants. 

In section \ref{sec:massive} we present the motivation to study the merger of two cores. In section \ref{sec:LCEE} we estimate the luminosity and the duration of bright ILOTs that might result from the merger of two massive giants, and in section \ref{sec:evolution} we examine the properties of two red supergiants binary systems that might merge.
We summarise our results in section \ref{sec:summary}.

% ==========================================================
\section{The case for merging giant stars}
 \label{sec:massive}
% ==========================================================

From stellar evolution calculations (e.g., \citealt{Ekstrometal2012, Choietal2016}) we find the following relation for the life span of a star of mass $M_i$ on the main sequence
\begin{equation}
\log \left( \frac{\tau_{\rm MS}}{10^{10} \yr} \right) \simeq  0.75(\log M_i)^2 - 3.3 \log M_i ,
\label{eq:Taums1}
\end{equation}
where mass is in solar units.
From that relation we find
\begin{equation}
\frac{d \log \tau_{\rm MS} }{d \log M_i} \simeq 1.5 \log M_i - 3.3.
\label{eq:Taums2}
\end{equation}
The total duration of the giant phases is $\tau_{\rm G} \simeq 0.1 \tau_{\rm MS}$. The condition for the two stellar giant phases to overlap reads
\begin{equation}
\frac{\Delta M}{M} \la 0.1  \left( - \frac{d \log \tau_{\rm MS} }{d \log M_i} \right)^{-1} ,
\label{eq:DeltaM}
\end{equation}
where $\Delta M = M_{1i} - M_{2i}$ is the difference between the initial masses of the two stars.
For a solar mass star this reads $(\Delta M/M)_1 \la 0.03$, while for two stars with $M_i \simeq 30 M_\odot$ this reads $(\Delta M/M)_{30} \la 0.09$. Consider that the fraction of binary massive stars is larger than that of low mass stars (e.g., \citealt{MoeDiStefano2017}), the merger rate of two massive giant stars relative to the total number of massive stars is significantly larger than that of two low mass stars. In both cases, though, the merger rate of two giants is non-negligible compared with the total merger rate at the specific mass range.

{{{{ The chance of any two stars to merger depends mainly on the orbital separation of the binary system being smaller than a critical value ($\approx {\rm few} -30 \AU$, depending on masses and eccentricity), and on the eccentricity (higher eccentricity increases the merger probability as periastron distance is shorter hence tidal interaction is stronger). For the specific case we study here time is critical also, as we want the giant phases of the two stars to overlap, and to make them merge during that overlapping time period. }}}}

The chance to merge {{{{ as two giants }}}}  is much larger before core helium exhaustion as the giants spend a longer time during that phase, {{{{ i.e., a larger overlapping giant phases time period. }}}} In particular for low mass stars, the red giant branch lasts much longer than the asymptotic giant branch.
We also note that rotation changes somewhat the evolution time. If the two stars have different rotation velocities this can increase or decrease somewhat the allowed mass difference.

One big difference between the present case and that of a compact companion that enters the giant envelope is that a giant companion cannot launch jets. Jets might help in removing the envelope \citep{ShiberSoker2018}. We consider therefore the case that the two cores spiral toward each other and reach a close distance, and even merge.

% ==========================================================
\section{The CEE luminosity}
 \label{sec:LCEE}
% ==========================================================

% =======================
\subsection{Photon-diffusion-dominated ILOT}
\label{subsec:energy}
% =======================

We consider here the spiralling-in process before the two cores might merge inside the common envelope.
{{{{ At this phase the system is basically composed of three components. The first is the extended-tenuous common envelope that is the merger of the two envelopes, the second and third components are the cores of the two giants. The entire system rotates around the center of mass but not as a solid body. Due to gravitational interaction of the two cores with the envelope they spiral-in toward each other and their orbital period decreases.}}}}

{{{{ The outer part of the envelope, outside the orbit of the two cores, cannot keep in pace with the two cores, and its angular velocity is lower. The two cores therefore transfer orbital angular momentum to the outer envelope and continue to spiral-in toward each other. The key issue when we have two cores that burn hydrogen and/or helium on their outskirts is that they cannot accrete mass from the common envelope at a high rate. Their accretion rate is limited to the very slow rate of nuclear burning  (the same as the core of a giant star accretes from its own envelope in single-star evolution). As the cores do not accrete mass, they do not launch jets that in some other cases of CEE might facilitate envelope removal (e.g., \citealt{Shiberetal2019}). This implies that there is no extra energy source of mass accretion onto the compact object. }}}}

{{{{ Therefore, the energy that is available to remove the common envelope is only the orbital gravitational energy that the two spiralling-in cores release. }}}} Most of this orbital gravitational energy is channelled to remove the envelope and accelerate it to its terminal velocity. The recombination energy of the hydrogen and helium in the expanding envelope does not contribute much to envelope removal as this energy is mainly radiated away and mostly adds to the  luminosity of the system, as we discuss below.  

We emphasise that we expect no jets during the main CEE, namely before the two cores merge. Therefore, the outflow from the ILOT will not be as asymmetrical as ILOTs that have jets. In the later the jets shape bipolar nebulae, such as some bipolar planetary nebulae that ILOTs might form, and the nebula of Eta Carinae (the Homunculus).
Hence, we consider the following spherically-symmetric treatment to be adequate for our purposes.   

{{{{ We first consider the case where a large fraction, but not all, of the common envelope is removed on a time scale shorter than the photon diffusion time that we derive below. In section \ref{subsec:Dust} we discuss the opposite inequality.  }}}}

Let a substantial fraction of the gas of the two merged envelopes (the common envelope), a mass of $M_{\rm CE}$, leave the system during the first phase of the CEE with a terminal velocity of $v_t$. (The rest of the envelope is lost on a longer time scale, hence with a much lower luminosity.) This gas becomes transparent for radiation diffusion on a time scale of (e.g., see discussion of the diffusion time for supernovae by \citealt{KasenWoosley2009})
\begin{equation}
t_{\rm diff} \simeq \frac{\kappa M_{\rm CE} }{4 c R_{\rm diff}},
\label{eq:tdiff1}
\end{equation}
where $\kappa$ is the opacity and $R_{\rm diff} = t_{\rm diff} v_t$.
Substituting scaled values we find
\begin{equation}
\begin{split}
t_{\rm diff} & \simeq 4  %4.08
\left(\frac{\kappa}{1 \cm^2 \g^{-1}} \right)^{1/2}
\\
&
\times
\left(\frac{M_{\rm CE}} {10 M_\odot} \right)^{1/2}
\left(\frac{v_t}{100 \km \s^{-1}} \right)^{-1/2} \yr.
\end{split}
\label{eq:tdiff2}
\end{equation}
If all the luminosity comes from recombination energy of solar composition, $E_{\rm rec}=3\times10^{46} (M_{\rm CE}/M_\odot) \erg$, we find the average recombination luminosity to be
\begin{equation}
\begin{split}
L_{\rm rec, diff} \approx \frac{E_{\rm rec}}{t_{\rm diff}} & \simeq  6 \times 10^5    %
\left(\frac{\kappa}{1 \cm^2 \g^{-1}} \right)^{-1/2}
\\
&
\times
\left(\frac{M_{\rm CE}} {10 M_\odot} \right)^{1/2}
\left(\frac{v_t}{100 \km \s^{-1}} \right)^{1/2} L_\odot.
\end{split}
\label{eq:Lrec}
\end{equation}
For the same values the kinetic energy is
\begin{equation}
E_{\rm kin} = 10^{48} \left(\frac{M_{\rm CE}} {10 M_\odot} \right)
\left(\frac{v_t}{100 \km \s^{-1}} \right)^{2} \erg \simeq 3.4 E_{\rm rec}.
%% Erec=2.95 e47erg.
\label{eq:Ekin}
\end{equation}

This recombination luminosity might be observed as a brightness that lasts for about few months to severeal years. 
We expect though, that the ejected mass will collide with previously ejected mass, or that during the late CEE phases mass that is ejected at higher velocities collides with earlier ejected mass. The collision transfers kinetic energy to luminosity. Overall, we expect to have a transient event, an ILOT, that lasts for several months to several years, with a luminosity of
  $L_{\rm ILOT} \approx {\rm few} \times 10^5 L_\odot - {\rm few} \times 10^6 L_\odot \simeq 10^{39} - 10^{40} \erg \s^{-1}$.

At an age of few months to a few years the outer radius of the gas is at $\simeq 10^{14} -  10^{15} \cm$, and for the above luminosity values the black body temperature is $T_{\rm eff, diff} \approx 3000 \K$. This is a red transient, as most ILOTs are, and might even become a bright infrared ILOT. In a recent study \cite{Jencsonetal2019} report the observations of infrared bright ILOTs (although they do not term them ILOTs). We raise the possibility that some infrared bright ILOTs might be the merger processes of two giant stars.

% =======================
\subsection{Slow ILOT}
\label{subsec:Dust}
% =======================
 
{{{{ In the initial spiralling-in phase of a CEE, the so called plunge-in phase, the companion spirals deep into the envelope within about the dynamical scale of the common envelope (\citealt{Ivanovaetal2013b} for a review). During that phase the two spiralling-in cores in our case might remove a large portion of the common envelope.
For the two cores of the two giants the plunge-in time is $t_{\rm PI} \simeq {\rm several} \times {\rm yr}$. This time scale is of the order of the diffusion time scale $t_{\rm diff}$ as given by equation (\ref{eq:tdiff2}). This implies that we might have two basic situations, namely,  of $t_{\rm PI} \la t_{\rm diff}$  and  $t_{\rm PI} \ga t_{\rm diff}$. 
 For cases where the merging process is shorter than the diffusion time, $t_{\rm PI} \la t_{\rm diff}$, the analysis of section \ref{subsec:energy} holds.  }}}}

{{{{ When the merging processes is longer than the diffusion time, i.e., $t_{\rm PI} \ga t_{\rm diff}$, the ILOT duration is dictated by the merging processes rather than by the photon diffusion time. This might change the observed event by allowing dust formation and by converting more kinetic energy to radiation. }}}} 
  
{{{{ Consider a volume of gas that left the star at time $t^\prime$. At time $t > t^{\prime} + t_{diff}$ this volume of gas had time to cool by radiation (and it further cools by adiabatic expansion). Dust forms now in this outflowing gas. 
Consider then a small ejected mass from the common envelope, say $M_{\rm out} \simeq 0.1-1 M_\odot$, that flows out at a velocity of $v_t \simeq 100 \km \s^{-1}$. It cools within a year (eq. \ref{eq:tdiff2}). Therefore, within few years we might have a shell of mass of $\simeq 1 M_\odot-{\rm few} \times M_\odot$ at a distance of $r_{\rm D} \simeq 10^{15} \cm$. During these several years the average luminosity of the recombining envelope in this case of a slow spiralling-in process is (for the parameters we took) 
$L_{\rm rec, slow} < L_{\rm rec, diff}$, where $L_{\rm rec, diff} \approx {\rm few} \times 10^5-10^6 L_\odot$ by equation (\ref{eq:Lrec}). However, a new process might contribute to the luminosity. }}}}
    
{{{{ When the spiralling-in process continues it might eject mass at higher velocities. As the newly ejected mas collides with previously ejected mass it might convert some of the kinetic energy to radiation. As the kinetic energy is only few times that of recombination (eq. \ref{eq:Ekin}), this process will at most double the luminosity from recombination. We term this factor of increasing radiated energy by the collision of the expanding gas with itself $\eta$. Overall, if we take the available energy to be twice the recombination energy, i.e., $\eta=2$, the luminosity in the case where $t_{\rm PI} > t_{\rm diff}$  is 
\begin{equation}
L_{\rm slow} \approx 5\times 10^5 \left( \frac{\eta}{2} \right) \left(\frac{M_{\rm CE}} {10 M_\odot} \right)
\left(\frac{t_{\rm PI}}{10 \yr} \right)^{-1} L_\odot.
\label{eq:Lslow}
\end{equation}
}}}}

{{{{ We now need to consider the dusty expanding shell that once was the common envelope. Over a time period of $t_{\rm PI} \approx 5 -50 \yr$ the expanding envelope reaches a distance of $r_d \approx 3 \times 10^{15} (v_t/100 \km \s^{-1}) (t_{\rm PI}/10 \yr) \cm$. The optical depth of this dusty shell is 
\begin{equation}
\begin{split}
\tau_d \approx  &  500 
\left( \frac{\kappa_d} {3 \cm^2 \g^{-1}} \right)
\left( \frac{M_{\rm CE}} {10 M_\odot} \right) 
\\ & 
\left( \frac{v_t} {100 \km \s^{-1}} \right)^{-2} 
\left( \frac{t_{\rm PI}}{10 \yr} \right)^{-2} .
\label{eq:TauDLslow}
\end{split}
\end{equation}
The dust opacity is as in dense winds from giants, $\kappa_d=1-10 \cm^2 \g^{-1}$ (e.g., \citealt{Hofneretal2003}). 
}}}}

{{{{ The dusty shell will obscure the ILOT and the remnant of the merger for tens to hundreds of years, with the photosphere at $\approx r_d$. For the average luminosity as give by equation (\ref{eq:Lslow}) the black-body temperature of the photosphere of the dusty shell for the parameters as we use here, and for the relevant time period of $\simeq 5-50 \yr$ is $T_{\rm eff,slow} \approx 700 (t_{\rm PI}/10 \yr)^{-3/4} \K$. This slow ILOT lasts for a longer time than the photon diffusion-dominated ILOT that we studied in section \ref{subsec:energy}, it is fainter, its radius is larger, and hence it is much redder.  During the relevant time, depending on the duration of the  plunge-in phase $t_{\rm PI}$, in this case the ILOT is a bright source in the IR band of about $2 \mu m -10 \mu m$, rather than in the red or in the very near IR as the photon diffusion-dominated ILOT. }}}}

% =======================
\subsection{Merger of the two cores}
\label{subsec:CoreMErger}
% =======================

After most of the envelope, but not all, leaves the binary cores the two cores might spiral in further due to several effects. (1) Some mass that is left around the binary system (e.g., \citealt{KashiSoker2011, ChenPodsiadlowski2017}). (2) Tidal interaction between the two cores. (3) Further evolution that causes the cores to expand.

If the two cores merge, they liberate a gravitational energy of
\begin{equation}
\begin{split}
& E_{\rm merg}  \simeq 0.5 \frac {G M_{\rm core,1} M_{\rm core,2}}
{R_{\rm core,1}+R_{\rm core, 2}} = 4.7 \times 10^{49}
\\
&
\times
\left[  \frac{M_{\rm core,1} M_{\rm core,2}}{ (5 M_\odot)^2} \right]
\left( \frac{R_{\rm core,1}+R_{\rm core, 2}}{1 R_\odot} \right)^{-1} \erg.
\label{eq:ECmerg}
\end{split}
\end{equation}
A large fraction of the merger energy goes to eject the rest of the common envelope and to bring it to its terminal velocity as it escapes the system.
If some of the envelope stays bound, it absorbs energy and inflates. After one core is destroyed onto the other core, a large fraction of the merger energy goes to uplift the destructed core mass to form an envelope around the more massive core. Only a small fraction of the total merger energy adds up to the radiated energy. However, the merger process is rapid, times scale of hours to days, and might power the ejected mass in a time scale of weeks to many years after CEE as the merger products reaches thermal equilibrium. Furthermore, fast ejected mass can catch up with previously ejected mass and increase the luminosity as kinetic energy is channelled to radiation.

% ==========================================================
\section{Examining two close-mass binaries}
 \label{sec:evolution}
% ==========================================================

% =======================
\subsection{Numerical setup}
\label{subsec:Numerics}
% =======================
We use the \textsc{mesa} code (Modules for Experiments in Stellar Astrophysics, version 10398; \citealt{Paxtonetal2011, Paxtonetal2013, Paxtonetal2015, Paxtonetal2018}) to study the parameter range leading to the spiraling-in process.
We examine some properties of two binary systems.
One binary system with  zero age main sequence (ZAMS) masses of $M_{\rm ZAMS,1}=15.75 M_\odot$ and $M_{\rm ZAMS,2}=15 M_\odot$, and a second binary system with $M_{\rm ZAMS,1}=31.5 M_\odot$ and $M_{\rm ZAMS,2}=30 M_\odot$.
Both systems are with a ZAMS metallicity of $Z=0.019$ and an equatorial velocity of $v_{\rm rot,i}=100 \km \s^{-1}$.
Stellar winds are taken
as \cite{Vink2001} for  $T_\mathrm{eff} \ge 11\,000\,\mathrm{K}$ and \cite{deJager1988} for $T_\mathrm{eff} \le 10\,000\,\mathrm{K}$.

% =======================
\subsection{Stellar structures}
\label{subsec:Stellar}
% =======================

\begin{figure}
\centering
%\hspace*{-1.3cm}
%% \vspace*{-3.0cm}
\includegraphics[width=0.49\textwidth]{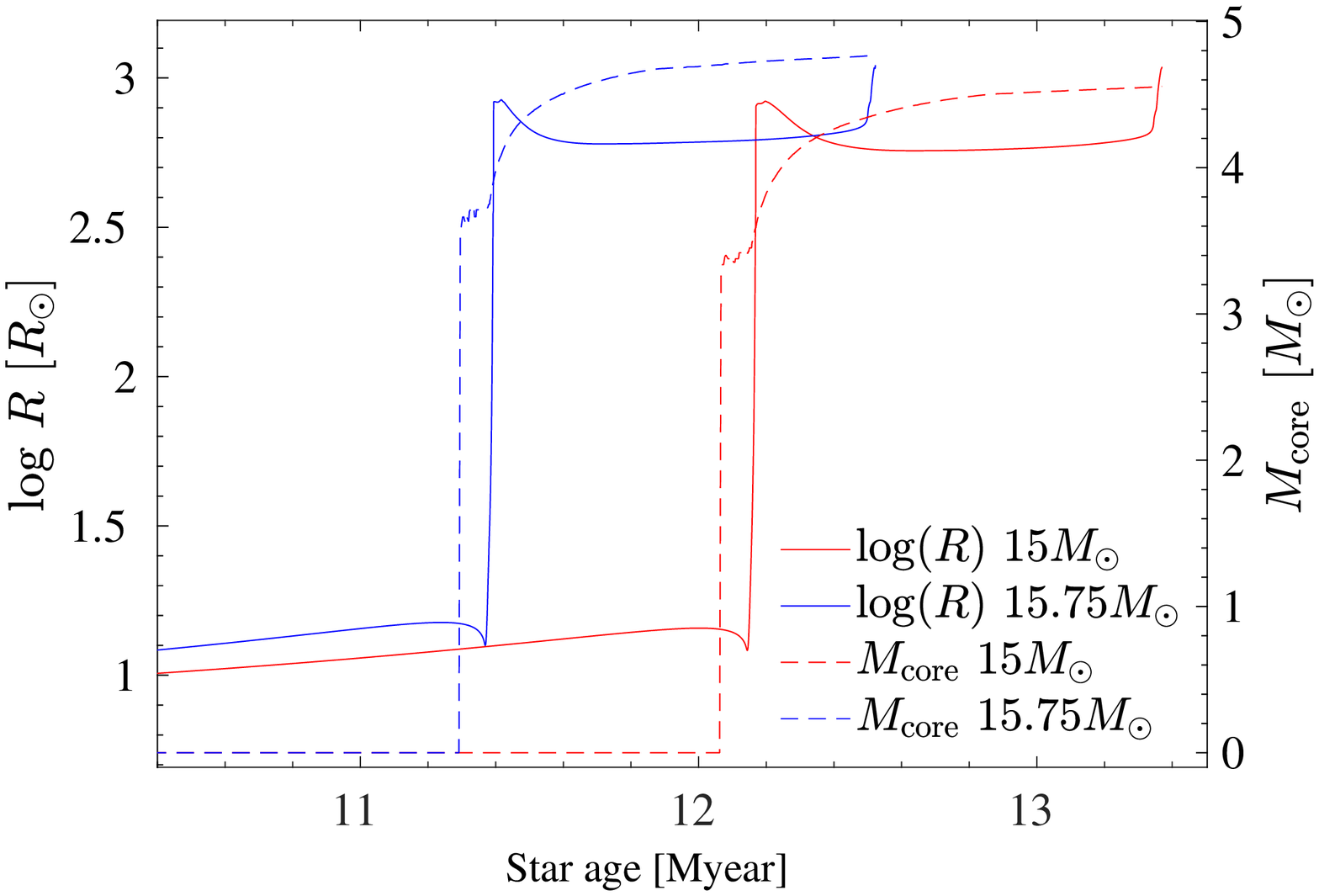}
%%% EXPLANATION:  trim={<left> <lower> <right> <upper>}
%%% \includegraphics[trim= 0cm 0.0cm 0cm 0cm, clip=true,  width=0.45\textwidth]{FigGiantMerger15Radii.eps}
%\hspace*{0.4cm}
%% \vspace*{-3.2cm}
\caption{The radii (solid lines) and core masses (dashed lines) of the two stars in the binary system, $M_{\rm ZAMS,1}=15.75 M_\odot$ (blue) and $M_{\rm ZAMS,2}=15 M_\odot$ (red) as function of time.
We assume that merger takes place when the secondary  experiences its large and rapid expansion. }
\label{Fig:Radii15Mo}
\end{figure}

\begin{figure}
\centering
%\hspace*{-1.3cm}
%% \vspace*{-3.0cm}
\includegraphics[width=0.49\textwidth]{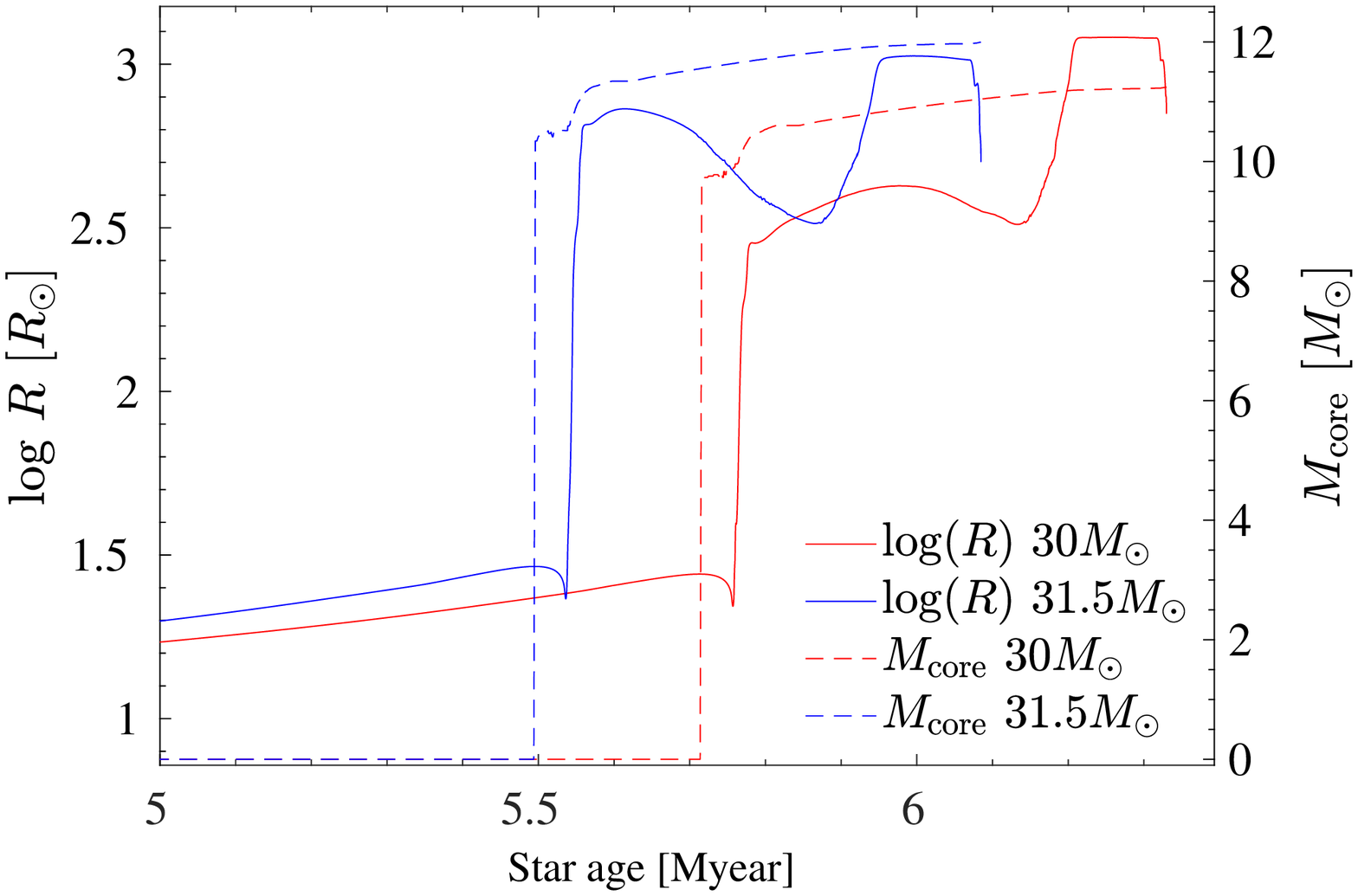}
%%% EXPLANATION:  trim={<left> <lower> <right> <upper>}
%%% \includegraphics[trim= 1cm 5.0cm 0.2cm 0.3cm,clip=true,width=0.55\textwidth]{FigGiantMerger30Radii.eps}
%\hspace*{0.4cm}
%% \vspace*{-3.2cm}
\caption{Like Fig. \ref{Fig:Radii15Mo} but for the $M_{\rm ZAMS,1}=31.5 M_\odot$  (blue) and $M_{\rm ZAMS,2}=30 M_\odot$ (red) binary system.  }
\label{Fig:Radii30Mo}
\end{figure}

We present the evolution of the radii and core masses of these two binary systems at late evolutionary phases in Figs. \ref{Fig:Radii15Mo} and \ref{Fig:Radii30Mo}, respectively. In the first binary the primary experiences its rapid expansion $7.8 \times10^5 \yr$  before the secondary star does, and in the second binary system the time difference is $2.2 \times10^5 \yr$.
We assume that the orbital separation between the two stars is such that the two giants merge when the secondary experiences its large expansion. We discuss later the case of mass transfer from the primary to the secondary star before the secondary expands.
% %% FFFFFFFFFFFFFFFFFFFFFFFFFFFFFFFFFFFFFFFFFFFFFFFFFF
% %% FFFFFFFFFFFFFFFFFFFFFFFFFFFFFFFFFFFFFFFFFFFFFFFFFF
% %% FFFFFFFFFFFFFFFFFFFFFFFFFFFFFFFFFFFFFFFFFFFFFFFFFF
% %% FFFFFFFFFFFFFFFFFFFFFFFFFFFFFFFFFFFFFFFFFFFFFFFFFF

\begin{figure}
%\vspace*{-1.0cm}
\centering
\includegraphics[width=0.41\textwidth]{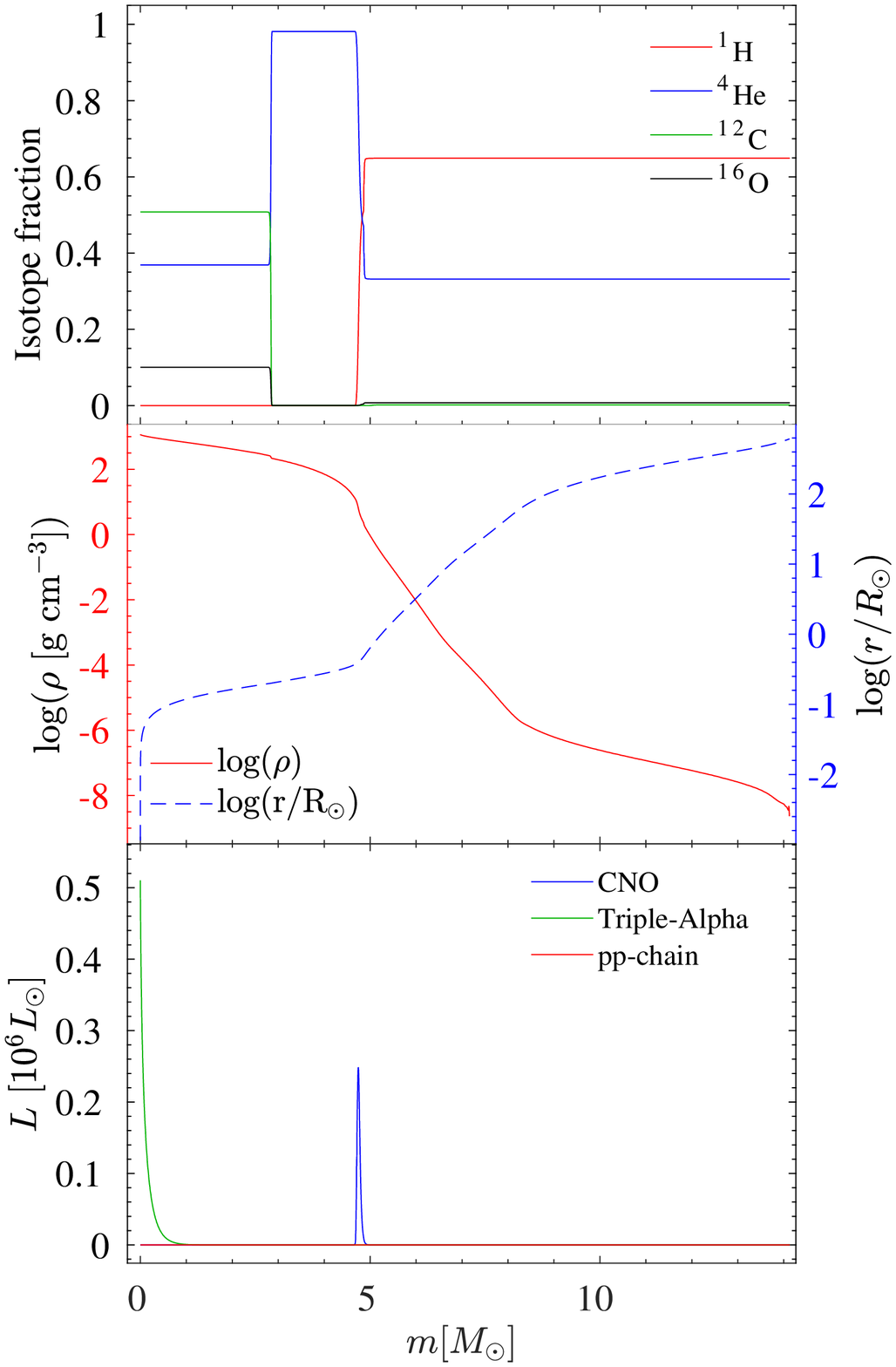}
\includegraphics[width=0.41\textwidth]{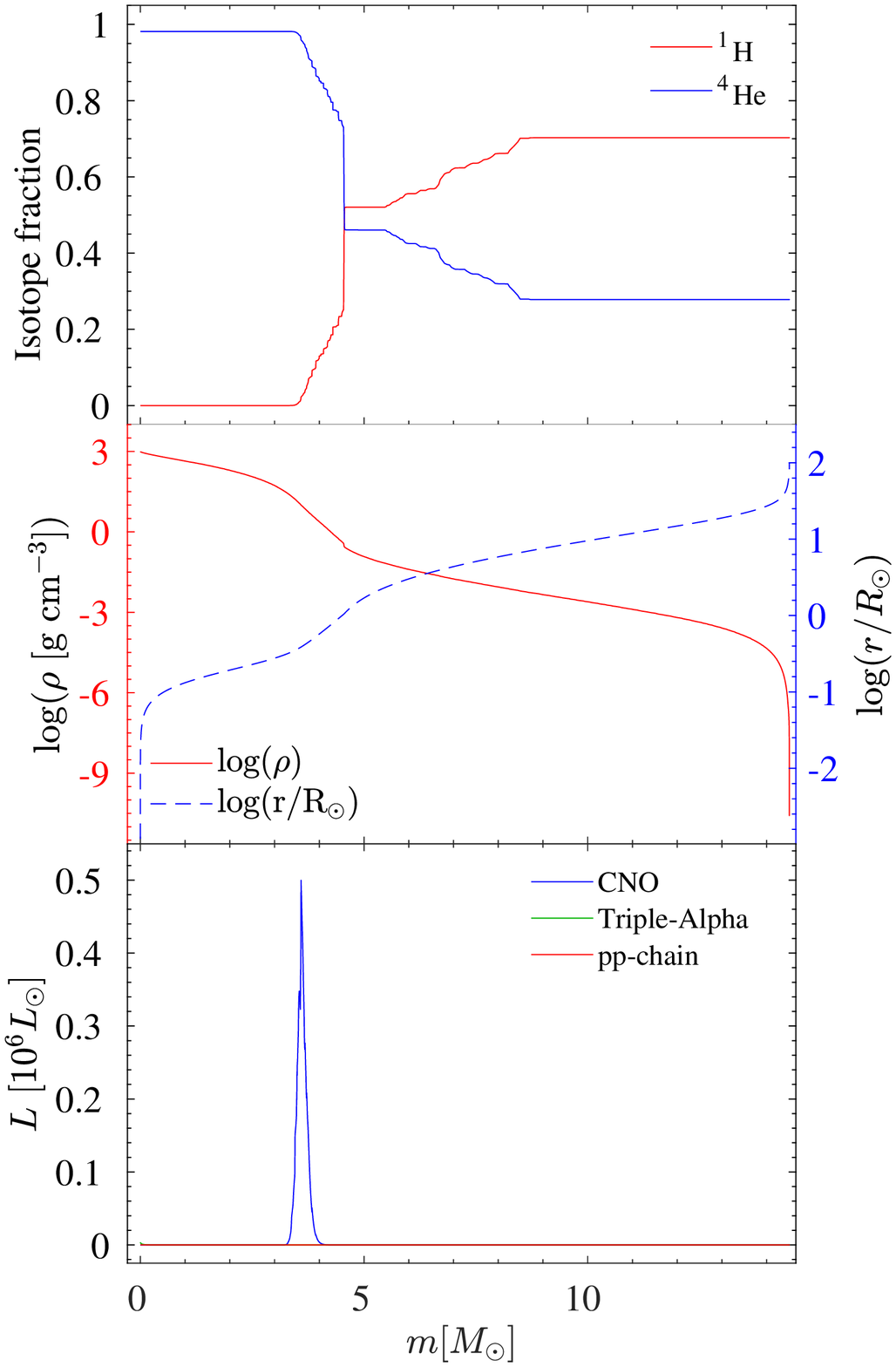}
% \hspace*{-1.3cm}
% \vspace*{-1.0cm}
% \includegraphics[trim= 3cm 2.5cm 2cm 3cm,clip=true,width=0.4\textwidth]{FigGiantMerger1575.eps}
% \hspace*{-1.5cm}
% \vspace*{-5.0cm}
% \includegraphics[trim= 3cm 2.5cm 2cm 3cm,clip=true,width=0.4\textwidth]{FigGiantMerger1500.eps}
%\hspace*{0.4cm}
% \vspace*{-1.cm}
\caption{The structures of the binary systems $M_{\rm ZAMS,1}=15.75 M_\odot$ (upper) and $M_{\rm ZAMS,2}=15 M_\odot$ (lower) when the lower mass companion experiences its large and rapid expansions at $t= 12.2 \times 10^6 \yr$.}
\label{fig:15Structures}
\end{figure}

\begin{figure}
\centering
%% \vspace*{-1.0cm}
\includegraphics[width=0.41\textwidth]{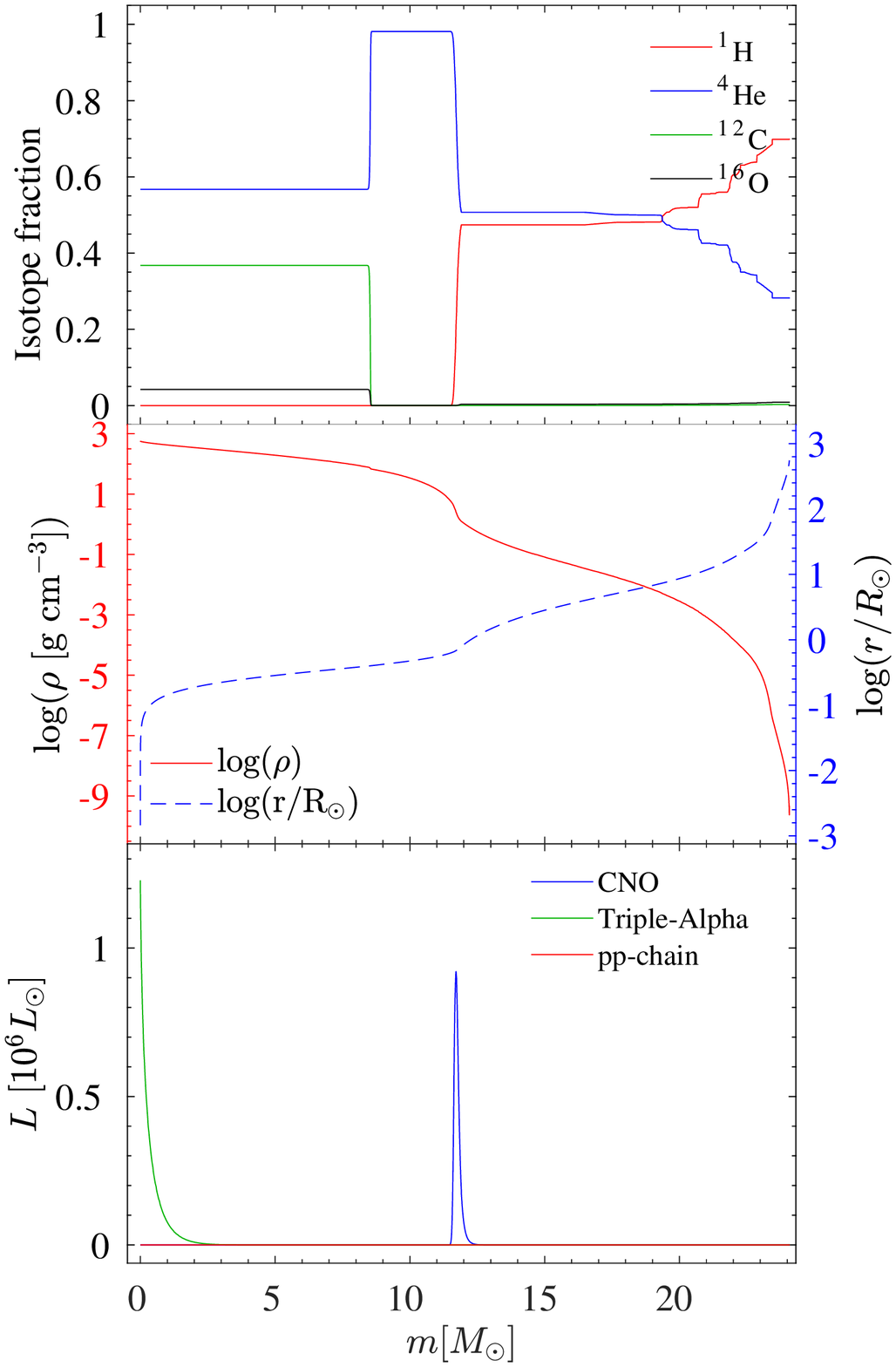}
\includegraphics[width=0.41\textwidth]{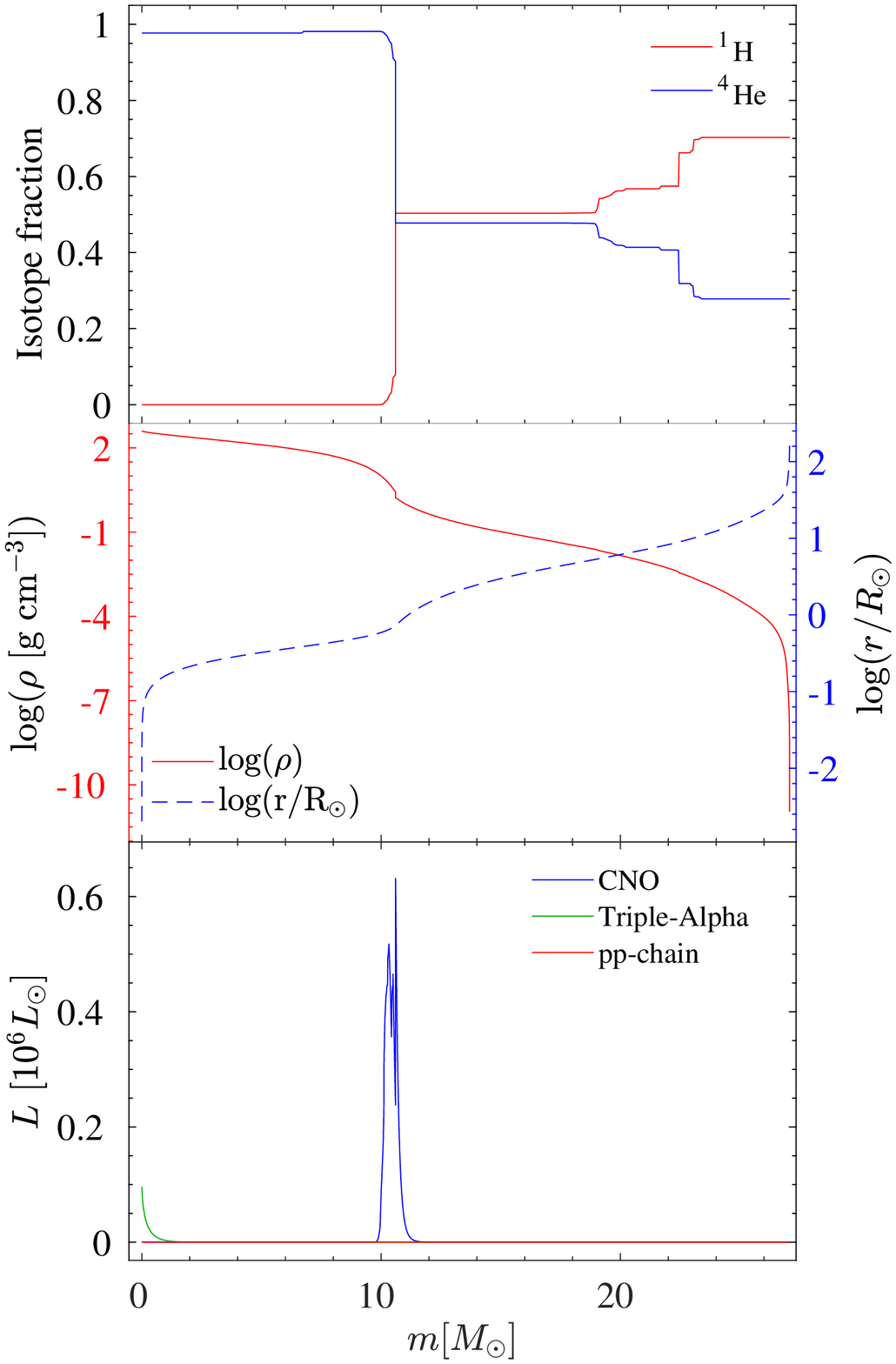}
%% \hspace*{-1.3cm}
%% \vspace*{-1.0cm}
% \includegraphics[trim= 3cm 2.5cm 2cm 3cm,clip=true,width=0.4\textwidth]{FigGiantMerger3159.eps}
%% \hspace*{-1.5cm}
% \includegraphics[trim= 3cm 2.5cm 2cm 3cm,clip=true,width=0.4\textwidth]{FigGiantMerger3000.eps}
%\hspace*{0.4cm}
%% \vspace*{-1.cm}
\caption{The structures of the stars in the binary systems $M_{\rm ZAMS,1}=31.5 M_\odot$ (upper) and $M_{\rm ZAMS,2}=30 M_\odot$ (lower) when the lower mass companion experiences its large and rapid expansion $t= 5.8 \times 10^6 \yr$.  }
\label{fig:30Structures}
\end{figure}

In Figs. \ref{fig:15Structures} and \ref{fig:30Structures} we present the stellar models of the stars in the two binary systems we study here when the lower mass companion experiences its large and rapid expansion. We assume that merger occurs during this phase.
% %% FFFFFFFFFFFFFFFFFFFFFFFFFFFFFFFFFFFFFFFFFFFFFFFFFF
% %% FFFFFFFFFFFFFFFFFFFFFFFFFFFFFFFFFFFFFFFFFFFFFFFFFF
% %% FFFFFFFFFFFFFFFFFFFFFFFFFFFFFFFFFFFFFFFFFFFFFFFFFF
% %% FFFFFFFFFFFFFFFFFFFFFFFFFFFFFFFFFFFFFFFFFFFFFFFFFF

In Table \ref{table:Properties1} we present the following stellar parameters for the two stars at the time when the more massive star (primary star) finishes its rapid expansion: luminosity $L$, effective temperature $T_{\rm eff}$, stellar masses $M_\ast$, core mass $M_{\rm core}$, stellar radius $R_\ast$,  core radius $R_{\rm core}$, and binding energy of the envelope $E_{\rm bind}$.
In Table \ref{table:Properties2} we present the same quantities but for the case when the secondary star finishes its rapid expansion. We assume that a merger takes place then (later we will discuss alternative merger times).
%TTTTTTTTTTTTTTTTTTTTTTTTTTTTTTTTTTTTTTTTTTTTTTTTTTTTTTTTTTTTTTTTTTTTTTTTTTTT
\begin{table}[H]
\begin{center}
 \begin{tabular}{|c c c c c c c c|}
 \hline
$M_{\rm ZAMS}$& $L$          & $T_{\rm eff}$& $M_\ast$  & $M_{\rm core}$ & $R_\ast$  & $R_{\rm core}$ & $E_{\rm bind}$ \\
$M_\odot$     & $10^4L_\odot$& $10^3K$      & $M_\odot$ & $M_\odot$      & $R_\odot$ & $R_\odot$      & $10^{49} \erg$ \\
%[0.5ex]
 \hline
 $15   $      & 3.89       &  24.82      &   $14.8$    & $-$             & 11        & $-$           & $-$ \\
 $15.75$      & 6.10       &  3.32       &   $15.2$    & $3.9$           & 745       & $0.35$       & $1.3$ \\
 $30   $      & 20.31      &  31.13      &   $28.3$    & $-$             & 15        & $-$           & $-$ \\
 $31   $      & 33.43      &  5.95       &   $28.0$    & $11.1$          & 546       & $0.61$       & $12.7$ \\
 \hline
\end{tabular}
\centering
\caption{ Properties of the giant primary star and the pre-giant secondary star when the more massive (primary) star finishes its rapid expansion. At this phase the core of the secondary star is not relevant for our study and we do not present its properties. }
\label{table:Properties1}
\end{center}
\end{table}
%TTTTTTTTTTTTTTTTTTTTTTTTTTTTTTTTTTTTTTTTTTTTTTTTTTTTTTTTTTTTTTTTTTTTTTTTTTTTTT
%TTTTTTTTTTTTTTTTTTTTTTTTTTTTTTTTTTTTTTTTTTTTTTTTTTTTTTTTTTTTTTTTTTTTTTTTTTTT
\begin{table}[H]
\begin{center}
 \begin{tabular}{|c c c c c c c c|}
 \hline
$M_{\rm ZAMS}$& $L$          & $T_{\rm eff}$& $M_\ast$  & $M_{\rm core}$ & $R_\ast$  & $R_{\rm core}$ & $E_{\rm bind}$ \\
$M_\odot$     & $10^4L_\odot$& $10^3K$      & $M_\odot$ & $M_\odot$      & $R_\odot$ & $R_\odot$      & $10^{49} \erg$ \\
%[0.5ex]
 \hline
 $15   $      & 6.83       &  3.26       &   $14.5$    & $3.73$         & 821       & $0.33$       & $1.1$ \\
 $15.75$      & 4.77       &  3.46       &   $14.1$    & $4.69$         & 610       & $0.38$       & $1.67$ \\
 $30   $      & 31.11      &  8.21       &   $27.0$    & $10.37$        & 276       & $0.59$       & $12.79$ \\
 $31   $      & 32.91      &  5.89       &   $24.1$    & $11.58$        & 552       & $0.64$       & $10.28$ \\
 \hline
\end{tabular}
\centering
\caption{Properties of the giant stars when the lower mass star in the binary system finishes its rapid expansion. We assume that in many cases merger takes place during this time (see Figs. \ref{Fig:Radii15Mo} and  \ref{Fig:Radii30Mo} for the models). }
\label{table:Properties2}
\end{center}
\end{table}
%TTTTTTTTTTTTTTTTTTTTTTTTTTTTTTTTTTTTTTTTTTTTTTTTTTTTTTTTTTTTTTTTTTTTTTTTTTTTTT

The gravitational energy that the two cores release when they merge, namely, when their separation is $a_f=R_{\rm core,1} + R_{\rm core,2}$, is given by equation (\ref{eq:ECmerg}).
The ratio of this energy to the binding energy of the two envelopes, $E_{\rm bind, envs} = E_{\rm bind,1} + E_{\rm bind,2}$ is
\begin{equation}
\begin{split}
\frac{E_{\rm merg}}{E_{\rm bind, envs}} &
\approx 1.5 % 1.44
\left(  \frac{ R_{\rm core,1} + R_{\rm core,2} } { 0.7 R_\odot }  \right)^{-1}
\left[  \frac{M_{\rm core,1} M_{\rm core,2}}{ (4 M_\odot)^2} \right]
\\& \times
\left(  \frac{E_{\rm bind,1} + E_{\rm bind,2}} { 3 \times 10^{49} \erg } \right)^{-1}.
\label{eq:Ecores}
\end{split}
\end{equation}
This is the typical ratio for the $M_{\rm ZAMS,1}=15.75 M_\odot$ and $M_{\rm ZAMS,2}=15 M_\odot$ binary system (using the data from Table \ref{table:Properties2}). For the $M_{\rm ZAMS,1}=31.5 M_\odot$ and $M_{\rm ZAMS,2}=30 M_\odot$ binary system this ratio is $0.8$ (using the data from Table \ref{table:Properties2}).
Since the efficiency of channelling the gravitational energy of the merging cores to envelope removal is $\alpha_{\rm CE} <1$, when the cores merge there is still bound envelope, but a large fraction of the envelope is leaving the system and a fraction might form an extended envelope.

% =======================
\subsection{The ILOT properties}
\label{subsec:ILOTproperties}
% =======================

Using these results and equation (\ref{eq:ECmerg}), we can crudely estiamte the total radiated energy in the ILOT.
Very crudely, in many ILOTs the  typical ratio of radiated to kinetic energy is $E_{\rm rad} \approx 0.1 E_{\rm kin}$ (e.g., \citealt{KashiSoker2010b}).
We take about half of the envelope to be ejected at the escape speed $V_{\rm eject} \simeq 100 \km \s^{-1}$, and the radiation to carry a fraction of $\chi \simeq 0.1$ of that energy. For an ejected mass of $ M_{\rm eject} = 10 M_\odot$
we find $E_{\rm rad} \approx 10^{47} \erg$. This is the general median value of the radiated energy of ILOTs (e.g., \citealt{Kashi2018}).

The photon diffusion time from an ejected mass of $ M_{\rm eject} = 10 M_\odot$ expanding at $V_{\rm eject} \simeq 100 \km \s^{-1}$, and with an average opacity of $\kappa \approx 1 \cm^2 \g^{-1}$ is about few years. This can be of the order of the merger process from an orbital separation of several astronomical units.
Over all, the merger of the two cores can have the properties of an ILOT lasting few years and radiating with a luminosity of $L \approx {\rm few} \times 10^5 L_\odot - 10^6 L_\odot$. This is about equal to the luminosity from the recombination as given in equation (\ref{eq:Lrec}). Adding together recombination and core merger, the ILOT of this fatal CEE lasts for about months to several years (or even up to few tens of years; see section \ref{subsec:Dust}) with a luminosity of $L_{\rm rad, tot} \approx 10^6 L_\odot$.

These properties make a brighter and longer ILOT than OGLE-2002-BLG-360, whose progenitor was a lower mass star at earlier evolutionary stages  \citep{Tylendaetal2013}. Hence, this was not a merger of two giants.
The common property of that merger and the scenario we study here is that they are both binary systems experiencing a long CEE that ended with a merger of the core with a compact companion, i.e., a fatal CEE. This was already suggested by \cite{Tylendaetal2013}. They also find that this system contains a massive dusty outflow.

The ILOT properties we estimate here are more like those of the luminous blue variable star R71 in the Large Magellanic Cloud \citep{Mehneretal2013}, but the progenitor of R71 had a radius of only $\simeq 100 R_\odot$ \citep{Mehneretal2017}.
The merger of two giants we consider here will be redder, and might even be detected only in the infrared.

We could not find an ILOT that fits our expectation from two merging giants. This is not in contradiction with our expectation as these events are very rare \citep{Soker2019Fatal}.

% =======================
\subsection{Mass transfer}
\label{subsec:MassTransfer}
% =======================

The initial orbital separation can be smaller than the maximum radius that the initial more massive (primary) achieves. In that case the more massive star transfers mass to the companion as it expands. \cite{VignaGomezetal2019} showed that even in this case the two stars can merge after they both have a hydrogen-exhausted core.
Because the two stars are of about equal masses, the secondary star brings the envelope of the expanding primary giant to synchronisation, i.e., between the spin and orbital period. This makes the tidal forces very small, and the secondary star does not spiral-in into the envelope of the giant star. Instead, the primary star transfers mass to the secondary star, and after the secondary becomes the more massive star, further mass transfer increases somewhat the orbital separation. The system will merge only after the secondary star becomes a giant.

The mass transfer from the primary to the secondary does not change much our conclusion for the present paper, because by the time the primary star suffers its rapid expansion its core is already 85 per cent or more of its final mass (Figs. \ref{Fig:Radii15Mo} and \ref{Fig:Radii30Mo}).
When the secondary expands and the primary core spirals inside its envelope, we have a CEE with core masses about the same as in cases without a mass transfer, and it does not matter much where the common envelope comes from. The energy that the two merging stars and spiralling-in cores release and the time scale of the process are  about the same as without mass transfer.
Despite our expectation that the implication of mass transfer for the ILOT will be small, there is a need for a more detailed simulation of evolution that includes mass transfer.

% =======================
\subsection{Rate of events}
\label{subsec:RateEvents}
% =======================

\cite{VignaGomezetal2019} study the merger of two post main sequence very massive stars, i.e., with ZAMS mass of $\ga 45 M_\odot$, as they are looking for a merger product that might lead to pair instability supernovae. They  estimate that between 13\% and 47\% of this type of close mass binaries are in the right orbital separation for them to merge while having each a helium core.  They further estimate that the merger rate of such two very massive stars that might lead to pair instability supernovae is in the range of about $10^{-5} -  0.003$ times the rate of core collapse supernovae (CCSNe). Since we deal here with lower mass stars, the rate of ILOTs of two merging giants is larger. \cite{Soker2019Fatal} crudely estimates that the rate of merger of two giants that lead to a CCSN is $\approx 0.01$ the rate of all CCSNe.

%%% The ILOT scenario works also for lower mass stars that do not end their lives as CCSNe. This is the case if the combined mass of the two giant is $\la 8 M_\odot$. However for lower mass stars, each of mass $\la 4 M_\odot$, the probability for a binary system with the right properties out of all stars with the same mass is lower than for more massive stars  (section \ref{sec:massive}). Therefore, the relative event rate of two giants merger out of all stars of the given mass decreases with decreasing stellar mass. There is a need for more thorough analysis, i.e., a population synthesis, to estimate the merger rates of two giants of low mass stars that form fainter ILOTs. Hence, although the number of lower mass stars, with initial mass in the range of $\simeq 1-4 M_\odot$, is larger than that of more massive stars, their contribution to the two giants merger scenario of ILOTs is not as large. We very crudely estimate the contribution of low mass stars whose merger product does not explode as a CCSN to be about equal to that of the more massive stars.

{{{{ To crudely estimate the number of ILOTs coming from merging giants we take the following stellar binary properties from \cite{Zapartasetal2017}, who also discuss some uncertainties. The initial mass function $dN \propto M^{-2.3} dM$, an orbital separation that is flat in log scale in the period range of $0.15< \log (P/{\rm day}) <3.5$, a mass ratio in binary stars that is flat in the range $0.1< M_2/M_1 <1$, i.e., $0< \Delta M/M < \Delta_m = 0.9$, and a binary fraction in the above period range of about $f_b \approx 50 \%$. We crudely approximate equation \ref{eq:DeltaM} for the condition of overlapping giant phases as ${\Delta M}/{M} \la 0.03 + 0.002 M \equiv \Delta_{\rm OG}$, where mass is in solar units. The fraction of systems that are binary systems where both stellar components have overlapping giant phases is then $f_{\rm OG, all} \approx (\int dN)^{-1} f_b \int (\Delta_{\rm OG}/\Delta_m) dN \approx 0.02$,  % 0.0202
where the integration is from $M_1=1 M_\odot$ to large masses and we substituted the above values for the different quantities in the integration. $f_{\rm OG, all}$ is the approximate fraction of all binary systems that have overlapping giants phases, whether they explode as CCSNe or form planetary nebulae, and whether they merge or not.  }}}}

{{{{ The large uncertainty regarding the merging probability is in the orbital separation that might lead the two stars to merge while both are giants. We very crudely take this range to be a factor of 2 in orbital separation, e.g., the initial orbital separation should be $a_{\rm min}=2 \AU \la a \la 4 AU = a_{\rm max}$. This gives a range of a factor of $2^{3/2}$ in orbital period, i.e., $\log (P_{\rm max}/P_{\rm min})=\log 2^{3/2} = 0.45$, and so the fraction of binary systems in the right orbital range is $f_a \approx [0.45/(3.5-0.15)]=0.13$.}}}}

{{{{ Therefore, the fraction of binary systems that form ILOTs by the merger of two giants is very crudely $f_{\rm G,ILOT} \approx  f_a f_{\rm OG,all} \approx 0.003$. %0.0027
This is the fraction from all systems (including single stars) with $M>1 M_\odot$. As the number of CCSNe is $f_{\rm CCSNe} \approx 0.1$ of these systems, we find the ratio of ILOTs from the merger of two giants to the number of CCSNe to be $f_{\rm G,ILOT}/f_{\rm CCSNe} \approx 0.03$. }}}} 

{{{{ The contribution from binaries that start with a primary stellar mass of $>8 M_\odot$ out of all giant binary mergers is $\approx 12 \%$ of the $f_{\rm G,ILOT}$. However, the merger of two stars of masses $\ga 4 M_\odot$ might lead to a massive core that explodes also as a CCSNe. The contribution from binary systems that start with a primary stellar mass of $>4 M_\odot$ out of all giant binary mergers is $\approx 25 \%$, or a fraction of $\approx0.0075$ of the number of CCSNe. This is close to the value of $\approx 0.01$ that \cite{Soker2019Fatal} estimated for the fraction of ILOTs from the merger of giants that will explode as CCSNe out of all CCSNe (keeping in mind the large uncertainties). 
}}}}

Overall, we crudely estimate the event rate of ILOTs coming from two merging giants to be $\approx 0.03$ times the event rate of CCSNe, but the uncertainties are large and the number can be in the range of $\approx 0.01-0.05$. The uncertainties come from uncertainties in the mass transfer physics and tidal interaction of two evolved stars, as well as from the initial parameters of binary systems.

% ==========================================================
\section{SUMMARY}
 \label{sec:summary}
% ==========================================================

We examined the possible observational signature of the merger of two giant stars in a binary system, which we suggest leads to a new type of a luminous transient (ILOT). The main properties and evolutionary phases of this novel type of ILOTs are as follows. 
    
The orbital separation between the two stars is such that they enter a CEE where the two cores spiral-in toward each other and eject a large fraction of the common envelope, or even all of it.
In cases where the two cores eject the entire envelope before they merge, they might survive.
For the two stars to simultaneously be in their giant phases, the two stars must have close masses on their ZAMS (section \ref{sec:massive}; eq. \ref{eq:DeltaM}).
If the two cores do not merge they might end up as two white dwarfs, or two neutron stars. In both cases a later merger is possible. We did not study these later evolutionary phases here.

In cases where the two cores do not merge, there are two energy sources to power the radiation of the ILOT. The first is the recombination energy of the ejected envelope. {{{{ Either the photon diffusion time determines the duration of the ILOT (eq. \ref{eq:tdiff2}; section \ref{subsec:energy}) leading to a luminosity of $\approx 10^6 L_\odot$ depending on the envelope mass (eq. \ref{eq:Lrec}), or the rapid spiralling-in phase determines the duration of the ILOT (section \ref{subsec:Dust}). }}}} The other energy source is the gravitational energy of the  spiralling-in cores that accelerates the outflowing envelope (eq. \ref{eq:Ekin}), even to a full merger of the cores (section \ref{subsec:CoreMErger}). If fast outflowing gas collides with earlier ejected slower  envelope gas, the collision transfers kinetic energy to thermal energy and radiation (section \ref{subsec:Dust}).

Using the stellar evolution numerical code MESA we evolved stars of two binary systems as we depict in Figs. \ref{Fig:Radii15Mo}-\ref{fig:30Structures}. In this study we evolved each of the stars as a single star. The next step will be to conduct a thorough study that includes the evolution of the two stars as a binary system, including tidal forces and mass transfer, e.g., by using the MESA binary code. This type of calculations has several free parameters and deserves its own study.
Although we have evolved only massive stars that end as CCSNe, the merger of two giants can take place in stars as low as $\simeq 1M_\odot$.

From the properties of the stars we found that the merger of the two cores releases gravitational energy that marginally ejects the entire envelope (eq. \ref{eq:Ecores}). This implies that in many cases the two cores merge, i.e., a fatal CEE, leading to a more massive core that in cases of massive stars later explodes as a CCSN \citep{Soker2019Fatal}. The merger of the two cores releases an amount of gravitational energy much larger than that of the recombination energy. However, most of this energy goes to unbind the envelope and accelerate it, as well as to inflate some of the mass of the destroyed core. The energy of the merging cores does not add much to the radiation energy.

The merger of two giant stars is different than that of a giant with a main sequence star in one important aspect. While a compact companion, such as a main sequence or a neutron star, can accrete mass from the giant envelope inside or outside the envelope (e.g., \citealt{Shiberetal2019}) and launch jets, the merger of two giants involves no jets, at least until the two cores merge.
The merger of the two cores can launch jets. The energy that such jets carry is part of the energy of the cores merger (equation \ref{eq:ECmerg}).
The jets can facilitate mass removal and power a brighter ILOT
(e.g., \citealt{SokerKashi2016}).
The descendant nebula that the out-flowing envelope forms is bipolar as the jets inflate two (or more if the jets precess) opposite bubbles. The merging two giants of the present study are not expected to form a bipolar nebula. They most likely form an elliptical nebula because the merger process ejects more mass in the equatorial plane.
  
The lack of jets, that implies that the outflow from the merger process of two giants does not deviate by a large degree from being spherically-symmetric, justifies our usage of spherically-symmetric photon diffusion treatment (section  \ref{sec:LCEE}).

Broadly speaking, the outcome of the  merger of two giant stars is a transient event, an ILOT, that lasts for several months to several years, with a luminosity of  $L_{\rm ILOT} \approx {\rm few} \times 10^5 L_\odot - {\rm few} \times 10^6 L_\odot \simeq 10^{39} - 10^{40} \erg \s^{-1}$, for a common envelope mass of $\approx 10 M_\odot$. The radiated energy scales more or less linearly with the common envelope mass. Due to high mass loss rate and massive outflow, we expect dust formation. The ILOT might be very red, or even infrared bright and undetectable in the visible. \cite{Jencsonetal2019} reported recently the observations of such infrared luminous transients, although most of these transients are likely to be infrared-bright and visibly-hidden CCSNe.

%\vspace{0.1cm}
% ==========================================================
\section*{Acknowledgments}
% ==========================================================
We thank Avishai Gilkis for his help in proper adjustment of the stellar parameters in MESA for the evolution of massive stars.
We thank Amit Kashi for useful comments.
{{{{ We thank an anonymous referee for suggestions to extend the discussion of some processes. }}}}
This research was supported by the Prof. A. Pazy Research Foundation and by a grant from the Israel Science Foundation.
%\vspace*{0.1cm}

\label{lastpage}
\end{document}